\documentclass[english]{paper}
\usepackage[T1]{fontenc}
\usepackage[latin9]{inputenc}
\usepackage[letterpaper]{geometry}
\usepackage{units}
\usepackage{amsthm}
\usepackage{amsmath}
\usepackage{esint}

\makeatletter

\newcommand{\noun}[1]{\textsc{#1}}

\numberwithin{equation}{section}
\numberwithin{figure}{section}

\usepackage{multicol}

\makeatother

\usepackage{babel}

\begin{document}
\begin{center}
\textbf{\huge On the logarithmic oscillator as a thermostat}
\par\end{center}{\huge \par}

\begin{center}
Marc Meléndez\\
\emph{\small Dpto. Física Fundamental, Universidad Nacional de
Educación a Distancia,}\\
\emph{\small Madrid, Spain}{\small .}
\par\end{center}{\small \par}
\begin{abstract}
Campisi, Zhan, Talkner and Hänggi have recently proposed \cite{Campisi}
the use of the logarithmic oscillator as an ideal Hamiltonian thermostat,
both in simulations and actual experiments. However, the system exhibits
several theoretical drawbacks which must be addressed if this thermostat
is to be implemented effectively.
\end{abstract}
\begin{multicols}{2}\thispagestyle{empty}

\section{The logarithmic oscillator}

A logarithmic oscillator is a point mass $m$ in a central logarithmic
potential. The Hamiltonian for such a particle is\begin{equation}
H_{osc.}\left(\boldsymbol{q},\,\boldsymbol{p}\right)=\frac{\boldsymbol{p}^{2}}{2m}+k_{B}T\,\ln\left(\frac{\left\Vert \boldsymbol{q}\right\Vert }{b}\right)=E,\label{eq:Hamiltonian-osc}\end{equation}
where $k_{B}T$ and $b$ can be considered arbitrary parameters for
the time being. The Hamiltonian equations of motion are therefore\begin{equation}
\left\{ \begin{array}{ccc}
\dot{q}_{i}=\frac{\partial H}{\partial p_{i}} & = & \frac{p_{i}}{m},\\
\dot{p}_{i}=-\frac{\partial H}{\partial q_{i}} & = & -k_{B}T\frac{q_{i}}{\boldsymbol{q}^{2}}.\end{array}\right.\label{eq:Eq_motion-osc}\end{equation}
This mechanical system has several interesting properties. 

In the one-dimensional version of the oscillator, it is particularly
easy to find the equations of motion by direct integration (we will
disregard the singularity in the potential for the moment). From \eqref{eq:Hamiltonian-osc},
we get the value of the momentum,\[
p=\sqrt{2m\left(E-k_{B}T\,\ln\left(\frac{q}{b}\right)\right)},\]
and using the first of Hamilton's equations of motion \eqref{eq:Eq_motion-osc},\[
\dot{q}=\sqrt{\frac{2}{m}\left(E-k_{B}T\,\ln\left(\frac{q}{b}\right)\right)},\]
we get a differential equation which can be solved by separation of
variables\begin{equation}
t=\sqrt{\frac{m}{2}}\int\frac{dq}{\sqrt{E-k_{B}T\,\ln\left(\frac{q}{b}\right)}}.\label{eq:t(q)}\end{equation}
Now, the amplitude of the oscillation is determined by the points
$q_{\alpha}$ that satisfy the following equation:\[
k_{B}T\,\ln\left(\frac{q_{\alpha}}{b}\right)=E,\]
that is,\begin{eqnarray*}
q_{A} & = & -be^{\beta E},\\
q_{B} & = & be^{\beta E},\end{eqnarray*}
where $\beta$ represents $\left(k_{B}T\right)^{-1}$. The period
of oscillation is just twice the time taken by the particle to go
from $q_{A}$ to $q_{B}$,\begin{equation}
2t_{AB}=\sqrt{2m}\int_{q_{A}}^{q_{B}}\frac{dx}{\sqrt{E-k_{B}T\,\ln\left(\frac{\left|q\right|}{b}\right)}}.\label{eq:2tAB}\end{equation}
The function in the integral is even, so\begin{eqnarray*}
2t_{AB} & = & \sqrt{8m}\int_{0}^{q_{B}}\frac{dx}{\sqrt{E-k_{B}T\,\ln\left(\frac{q}{b}\right)}}\\
 & = & \sqrt{\frac{8\pi m}{k_{B}T}}be^{\beta E}.\end{eqnarray*}

In the more general case, the motion of the particle lies on a plane.
If it moves in circular orbits around the singularity with a radius
$r$, then its velocity can be deduced from the fact that the central
and centrifugal forces must balance,\[
F=\frac{k_{B}T}{r}=m\frac{v^{2}}{r}.\]
Therefore, the speed\begin{equation}
v=\sqrt{\frac{k_{B}T}{m}}\label{eq:v-circular-orbits}\end{equation}
does not depend on the radius of the orbit. The radius of the orbit
is a function of the total energy $E$, because inserting \eqref{eq:v-circular-orbits}
into \eqref{eq:Hamiltonian-osc}, setting $q$ equal to $r$ and then
solving for $r$ gets us\[
r=\frac{be^{\beta E}}{\sqrt{e}}.\]
Therefore, the time it takes the particle to complete an orbit is\[
t_{orb.}=\frac{2\pi r}{v}=2\pi\sqrt{\frac{m}{ek_{B}T}}be^{\beta E}.\]

For arbitrary initial conditions, the trajectory followed by the oscillator
will not usually be a closed path, but the particle will never move
further out than\[
r_{max.}=be^{\beta E},\]
for a given energy $E$, and the time between two consecutive maximum
distances will be somewhere between $2t_{AB}$ and $t_{orb.}$ (note
that both times are of the same order of magnitude),\begin{equation}
\frac{2t_{AB}}{t_{orb.}}=\sqrt{\frac{2e}{\pi}}.\label{eq:magnitude-tper}\end{equation}

\section{Statistical properties}

The fact that the speed on a circular orbit does not depend on the
radius is quite surprising. It implies that, if an external perturbation
were to relocate the oscillator on a new circular orbit, the kinetic
energy would remain the same and all the energy absorbed would be
completely converted into potential energy.

In a sense, this result can be generalised to the oscillator's other
trajectories. If we define the virial $G$ as\begin{equation}
G=pr,\label{eq:virial-definition}\end{equation}
and calculate its time derivative using \eqref{eq:Eq_motion-osc},\[
\frac{dG}{dt}=p\dot{r}+\dot{p}r=2\left(\frac{p^{2}}{2m}\right)-k_{B}T.\]
The time average of the previous formula is\[
\left\langle \frac{dG}{dt}\right\rangle _{t}=2\left\langle \frac{p^{2}}{2m}\right\rangle _{t}-k_{B}T,\]
and if $\left\langle dG/dt\right\rangle _{t}=0$, then the average
kinetic energy must be\begin{equation}
\left\langle \frac{p^{2}}{2m}\right\rangle _{t}=\frac{1}{2}k_{B}T,\label{eq:average-kinetic-energy}\end{equation}
\emph{whatever the value of} $E$! This means that the logarithmic
oscillator can absorb an arbitrary amount of energy without changing
its temperature at all, behaving (in a way) like an ideal thermostat.

Is it true, then, that $\left\langle dG/dt\right\rangle _{t}=0$?
It certainly is, as\begin{eqnarray}
\left\langle \frac{dG}{dt}\right\rangle _{t} & = & \lim_{t\rightarrow\infty}\frac{1}{t}\int_{0}^{t}\frac{dG}{d\tau}d\tau\label{eq:dGdt-is-zero}\\
 & = & \lim_{t\rightarrow\infty}\frac{G\left(t\right)-G\left(0\right)}{t}=0,\nonumber \end{eqnarray}
because $G$ has upper and lower bounds, as one can see by noting
that $G$ is a continuous function, except at the origin. Given that\begin{eqnarray*}
\lim_{r\rightarrow0}G\left(r\right) & = & 0,\\
G\left(r_{max.}\right) & = & 0,\end{eqnarray*}
we can infer that $G\left(r\right)$ has upper and lower bounds in
the interval $\left(0,\, r_{max}\right]$, and \eqref{eq:average-kinetic-energy}
is correct. However, we must not forget that there is a limiting process
involved in \eqref{eq:dGdt-is-zero}, and hence it might take a very
long time for the average kinetic energy to converge to $k_{B}T/2$.
In fact, we will argue that this is generally the case, and that the
logarithmic oscillator is therefore a somewhat less-than-ideal thermostat.

A recent article in the ar$\chi$iv \cite{Campisi} argued that weak
coupling between a system of interest and a logarithmic oscillator
will result in canonical sampling of the former's phase space. The
dynamics of the compound system would then be determined by a total
Hamiltonian\begin{eqnarray*}
H\left(\boldsymbol{q},\,\boldsymbol{p},\, r,\, p_{r}\right) & = & H_{S}\left(\boldsymbol{q},\,\boldsymbol{p}\right)+H_{osc.}\left(r,\, p_{r}\right)\\
 &  & +H_{int.}\left(\boldsymbol{q},\,\boldsymbol{p},\, r,\, p_{r}\right)=E,\end{eqnarray*}
where $H_{S}\left(\boldsymbol{q},\,\boldsymbol{p}\right)$ is the
Hamiltonian for the system of interest, $H_{osc.}\left(r,\, p_{r}\right)$
is the one-dimensional version of \eqref{eq:Hamiltonian-osc}, and
$H_{int.}$ is the potential energy of the weak interaction between
the system and the oscillator, which we will assume is negligible
compared to $H_{S}$ and $H_{osc.}$. The density of states for the
logarithmic oscillator is\begin{eqnarray*}
\Omega_{osc.}\left(E_{osc.}\right) & = & \int\delta\left(H_{osc.}\left(r,\, p_{r}\right)-E_{osc.}\right)\, dp_{r}\, dr,\end{eqnarray*}
with $\delta$ representing the Dirac delta function. The integral
turns out to be exactly the same as \eqref{eq:2tAB}, so\begin{equation}
\Omega_{osc.}\left(E_{osc.}\right)=\sqrt{\frac{8\pi m}{k_{B}T}}be^{\beta E_{osc.}}.\label{eq:Omega(Eosc)}\end{equation}
Furthermore, the probability density $\rho$ for a point in the phase
space of the system corresponding to $H_{S}$ is\begin{equation}
\rho\left(\boldsymbol{q},\,\boldsymbol{p}\right)=\frac{\Omega_{osc.}\left(E-H_{S}\left(\boldsymbol{q},\,\boldsymbol{p}\right)\right)}{\Omega\left(E\right)}.\label{eq:rho(q,p)}\end{equation}
The function $\Omega\left(E\right)$ represents the density of states
of the compound system,\begin{equation}
\Omega\left(E\right)=\int\delta\left(E-H\left(\boldsymbol{q},\,\boldsymbol{p},\, r,\, p_{r}\right)\right)\, dr\, dp_{r}\, d\boldsymbol{q}\, d\boldsymbol{p}.\label{eq:Omega(E)}\end{equation}
Expressions \eqref{eq:Omega(Eosc)} and \eqref{eq:Omega(E)} can be
used to convert \eqref{eq:rho(q,p)} into\[
\rho\left(\boldsymbol{q},\,\boldsymbol{p}\right)=\frac{e^{-\beta H_{S}\left(\boldsymbol{q},\,\boldsymbol{p}\right)}}{\int e^{-\beta H_{S}\left(\boldsymbol{q},\,\boldsymbol{p}\right)}d\boldsymbol{q}\, d\boldsymbol{p}},\]
which is precisely the canonical distribution for $H_{S}$.

According to the authors of \cite{Campisi}, the logarithmic oscillator
thermostat has two obvious advantages. Firstly, contrary to the popular
Nosé-Hoover thermostat, the dynamical equations of motion are Hamiltonian.
Secondly, it is possible to design experimental setups in which the
thermostat is an actual \emph{physical} system. Hoover wrote a reply
\cite{Hoover} to the first claim arguing that Nosé-Hoover mechanics
\emph{are} in fact Hamiltonian, and included an example of an alternative
Hamiltonian thermostat of the Nosé-Hoover type. Campisi \emph{et alii}
answered explaining their claim further in \cite{Campisi2}. Here
we will be considering the second claim instead, that is, we will
concentrate on the implementation of the logarithmic oscillator as
a thermostat, both in experiments and simulations.

\section{Experiments}

An experimental thermostat that relies on the dynamics of only a few
degrees of freedom is no doubt a very interesting system. However,
the nature of the logarithmic oscillator imposes some serious limitations
which must be taken into account before one attempts to design such
an experiment.

The first problem is a consequence of the length-scales involved.
Assume that we wish to bring a system with $N$ degrees of freedom
to the equilibrium temperature $T$. If the kinetic energy per degree
of freedom is initially off by a fraction $\alpha$ of the energy,\[
\left\langle \frac{p_{i}^{2}}{2m}\right\rangle _{t}=\left(1+\alpha\right)\frac{1}{2}k_{B}T,\]
then the logarithmic oscillator will have to absorb at least an amount
of energy equal to $\Delta E=N\alpha k_{B}T/2$. We have seen that
the oscillator typically covers distances of the order of $b\,\exp\left\{ \beta E_{osc.}\right\} $.
The change in energy implies that the distances covered will change
by\begin{equation}
\Delta r_{max.}=r_{max.}\left(e^{\beta\Delta E}-1\right).\label{eq:delta-rmax}\end{equation}
This can be problematic if $r_{max.}$ is initially comparable to
the size of the experimental apparatus and the oscillator is cooling
the system.

The enormous changes in lengths imply similar changes in time scales.
Having assumed a weak interaction between the system of interest and
the oscillator, the effect of the interaction on the latter during
one period of oscillation should not be significant. The period is\[
t_{per.}=\lambda\sqrt{\frac{m}{k_{B}T}}be^{\beta E_{osc.}},\]
where $\lambda$ is a factor that depends on the trajectory, but which
is of the order of magnitude of $\sqrt{8\pi}$, in agreement with
\eqref{eq:magnitude-tper}. The change in distances carries with it
a corresponding change in periods of oscillation,\begin{equation}
\Delta t_{per.}=t_{per.}\left(e^{\beta\Delta E}-1\right).\label{eq:delta-tper}\end{equation}
Therefore, when the oscillator is cooling down the system of interest,
it will usually move very far out and oscillate very slowly. On the
other hand, when it is {}``hotter'' than the system, it will squeeze
into a small neighborhood of the singularity and vibrate very quickly.

Let us illustrate the problem with some numbers. The authors of \cite{Campisi}
propose an experiment in which a small system composed of neutral
atoms is contained in a box of length $L$. The logarithmic oscillator
is an ion in a two-dimensional Coulomb field generated by a charged
wire.

Assume, for example, that we have a dilute gas of $10$ atoms
of argon at an initial temperature $T_{0}=3\,\unit{K}$ and that we
wish to bring them to $T=1\,\unit{K}$. This means that the logarithmic
oscillator must absorb about\begin{equation}
\Delta E=\frac{3}{2}Nk_{B}T_{0}-\frac{3}{2}Nk_{B}T=30\, k_{B}T\label{eq:Delta-E}\end{equation}
units of energy. Let us assume further that the cross section of the
charged wire has a radius equal to $10^{-3}\, L$. Then the logarithmic
oscillator must move in orbits with\[
r_{max.}>10^{-3}\, L.\]
However, when we insert \eqref{eq:Delta-E} into \eqref{eq:delta-rmax}
we find that\[
\Delta r_{max.}=r_{max.}\left(e^{30}-1\right)>10^{10}\, L.\]
If we also take equation \eqref{eq:delta-tper} into account, it is
easy to see that we should expect to find the oscillator outside the
box most of the time.

\section{Simulations}

The wide range of time and length scales affects the precision and
time of computation of numerical simulations as well, but the presence
of a singularity in the logarithmic potential introduces another complication
in the numerical implementation of the oscillator, as stepping over
the singularity will usually lead to the wrong energy $E_{osc.}$.

When the particle is in the vicinity of the singularity, the slope
$\partial H/\partial r$ changes very quickly. If the oscillator ends
up too close to the singularity, it will feel a great force which
will push it away from the singularity during the next time step,
making it skip the area in which the potential would slow it down
again, unless a very small time step is chosen.

For the one-dimensional version of the logarithmic oscillator, the
problem can be solved by calculating the new position of the logarithmic
oscillator first. If the oscillator has stepped over the singularity,
then expression \eqref{eq:t(q)} can be used to calculate the time
it would have taken to get to the new position, and one can reset
its kinetic energy to the correct value and calculate the evolution
of the system of interest during that time. This solution is far from
satisfactory, though, because it involves finding numerical values
of the error function every time the particle passes the singularity.

A different approach (\cite{Campisi}) replaces the logarithmic potential
with the approximate potential\[
V\left(r\right)=\frac{1}{2}k_{B}T\,\ln\left(\frac{r^{2}+b^{2}}{b^{2}}\right),\]
thereby eliminating the singularity and introducing only a slight
correction in the density of states for low values of $E_{osc.}$.
Unfortunately, this imposes a limit on the amount of energy available
for exchange between the oscillator and the system. If the system
and oscillator are enclosed in a box of length $L$, one only has
about $k_{B}T\,\ln\left(L/b\right)$ units of energy to play with.
In order to allow for larger energy ranges, one must choose smaller
values of $b$ (of the order of $\exp\left\{ -2\alpha3N\right\} $
if we wish to allow the energy to fluctuate by a fraction $\alpha$
either way), and this will tend to generate a small neighbourhood
of $r=0$ in which the forces on the oscillator are huge.

\section{Conclusions}

The logarithmic oscillator proposed by Campisi, Zhan, Talkner and
Hänggi displays very interesting properties from the point of view
of theoretical statistical mechanics. However, before it can be used
as a thermostat in actual experiments and numerical simulations, three
problems must be addressed. Firstly, the distances covered by the
oscillator depend exponentially on its energy. Given that it must
not interact strongly with container walls or other objects, one would
expect that it would be very difficult to control such a system in
practice. Secondly, the vast increase in the period of oscillation
when a system is being cooled down suggests that the desired thermostated
dynamics will be achieved very slowly. Lastly, the presence of a singularity
introduces some technical complications in the numerical implementation
of the dynamical behaviour of the oscillator. It seems, therefore,
that Nosé-Hoover dynamics will remain a popular option in molecular
dynamics at least until the problems mentioned here are resolved satisfactorily.

\section*{Aknowledgments}

The author would like to express his gratitude to Pep Español for his helpful
comments.

\end{multicols}
\end{document}